\documentclass[pra,aps,twocolumn,showpacs,preprintnumbers,amsmath,amssymb,floatfix,superscriptaddress]{revtex4}
\usepackage{dcolumn}
\usepackage{bm}
\usepackage{amsmath,amssymb,amsfonts,latexsym,fancyhdr,graphicx}

\newcommand{\ket}[1]{\displaystyle{|#1\rangle}}

\newcommand{\g}{\gamma}
\newcommand{\si}{\sigma}
\newcommand{\la}{\lambda}

\begin{document}
\title{Phenomenological memory-kernel master equations and time-dependent Markovian processes}
\author{L. Mazzola}\email{laumaz@utu.fi}
\affiliation{Turku Centre for Quantum Physics, Department of Physics and Astronomy, University of
Turku, FI-20014 Turun yliopisto, Finland}
\author{E.-M. Laine}\email{emelai@utu.fi}
\affiliation{Turku Centre for Quantum Physics, Department of Physics and Astronomy, University of
Turku, FI-20014 Turun yliopisto, Finland}
\author{H.-P. Breuer}
\affiliation{Physikalisches Institut, Universit\"at Freiburg, Hermann-Herder-Strasse 3, D-79104 Freiburg, Germany}
\author{S. Maniscalco}
\affiliation{Turku Centre for Quantum Physics, Department of Physics and Astronomy, University of
Turku, FI-20014 Turun yliopisto, Finland}
\author{J. Piilo}
\affiliation{Turku Centre for Quantum Physics, Department of Physics and Astronomy, University of
Turku, FI-20014 Turun yliopisto, Finland}

\begin{abstract}
 Do phenomenological master equations with memory kernel always
describe a non-Markovian quantum dynamics characterized by reverse flow of
information?
 Is the integration over the past states of the
system an unmistakable signature of non-Markovianity? We show by a
counterexample that this is not always the case. We consider two
commonly used phenomenological integro-differential master
equations describing the dynamics of a spin 1/2 in a thermal bath.
By using a recently introduced measure to quantify
non-Markovianity [H.-P. Breuer, E.-M. Laine, and J. Piilo,
Phys.~Rev.~Lett.~\textbf{103}, 210401 (2009)] we demonstrate that
as far as the equations retain their physical sense, the key
feature of non-Markovian behavior does not appear in the
considered memory kernel master equations. Namely, there is no
reverse flow of information from the environment to the open
system. Therefore, the assumption that the integration over a
memory kernel always leads to a non-Markovian dynamics turns out
to be vulnerable to phenomenological approximations.  Instead, the considered phenomenological
equations are able to describe time-dependent and uni-directional information flow from the system
to the reservoir associated to time-dependent Markovian processes.
\end{abstract}
\pacs{03.65.Yz, 03.65.Ta, 42.50.Lc}

\maketitle

\section{Introduction}
The study of non-Markovian open quantum systems has attracted
extraordinary attention and efforts in recent years \cite{Breuer}.
Many analytical methods and numerical techniques have been
developed to treat non-Markovian processes
\cite{pseudomode,NMQJ,semi-Markov,Budini,Daffer,SabrinaCP,SabrinaP,Pascazio,ShabLid,Kossakowski}.
In addition to their importance in addressing fundamental
questions \cite{border}, this is mainly due to the applications
non-Markovian systems find in many branches of physics.
Non-Markovian processes appear in quantum
optics~\cite{Breuer,Gardiner96a, Lambro}, solid state
physics~\cite{SS}, quantum chemistry~\cite{QC}, quantum
information processing~\cite{QIP}, and even in the description of
biological systems~\cite{Rebentrost}.

Recently, several more rigorous definitions and quantifications of
non-Markovian behavior in open quantum systems have been proposed
\cite{Wolf,NMmeasure,NMmeasure2,Plenio,Fisher}. In fact, in the past the
concept of non-Markovian dynamics has been quite loosely
defined. The term non-Markovian process could, e.g., stand for:
Not describable by a master equation with Lindblad structure, or
leading to non-exponential decay, or characterized by a
time-dependent generator, or involving an integral over the past
states of the system.

Quite often it has been argued that the treatment of non-Markovian
dynamics necessarily requires solving an integro-differential
equation for the reduced density matrix of the system. However, it
has been shown that master equations which are local in time can
also represent the memory effects of a non-Markovian process (see,
e.g., Ref.~\cite{Breuer} and references therein), without the need
to take into account a time integration over the past history of
the system. Here we go a step forward: Not only that memory kernel
master equations are not the unique tool to treat non-Markovian
dynamics, but we also demonstrate that the presence of a memory
kernel alone does not guarantee the non-Markovian character of
the process associated to the reverse flow of information from the
system to the environment.
This surprising result is obtained by applying a
recently proposed measure of non-Markovianity \cite{NMmeasure,NMmeasure2} to
two quite commonly used non-local master equations: The
generalized memory kernel master equation discussed by one of us~\cite{SabrinaCP} and the
post-Markovian Shabani-Lidar master equation~\cite{ShabLid}, both
used to study the time-evolution of a spin 1/2 in a thermal bath.
Our results are connected to the issue of phenomenological
vs.~microscopically derived master equations in quantum optics
which do not always produce coinciding results in all of the
relevant parameter regimes~\cite{Scala}. Here we show that there
can be also qualitative differences in addition to the
quantitative ones when the two approaches are used.

Before proceeding with our treatment we would like to emphasize
that we do not intend to discourage from the use of memory kernel
master equations, which indeed in many cases constitute a
fundamental tool to study non-Markovian systems (see for example
Ref.~\cite{Daffer}). Instead, we rather want to point out that the
mathematical features of phenomenological kernel equations do not guarantee that
the feedback effects from the environment and the physical
characteristics of non-Markovian dynamics are taken into account.

We begin by briefly describing the measure for non-Markovianity
introduced by three of us in Refs.~\cite{NMmeasure,NMmeasure2}. Then we
present the master equations under investigation and their
solutions for the density matrix of the reduced system. Finally,
we use these solutions to evaluate the degree of non-Markovianity
for the two quantum processes, concluding with a discussion and
some remarks.

\section{Measure for non-Markovianity}

The construction of the measure for the degree of non-Markovianity
in open systems is based on the definition of Markovian
processes as those that continuously reduce the distinguishability of
quantum states \cite{NMmeasure}. One can interpret this loss of distinguishability
as a flow of information from the open system to its environment.
By contrast, in a non-Markovian process there exists a pair of
states the distinguishability of which grows for certain times.
This growth of the distinguishability of states can be interpreted
as a reverse flow of information from the environment to the open
system which is defined to be the essential feature of
non-Markovianity~\cite{NMmeasure,NMmeasure2}.

An appropriate measure for the distinguishability between two
quantum states given by density matrices $\rho_1$ and $\rho_2$ is
the trace distance \cite{Nielsen}
\begin{equation}\label{trace}
D(\rho_{1},\rho_{2})=\frac{1}{2}\mathrm{Tr}|\rho_{1}-\rho_{2}|,
\end{equation}
where $|A|=\sqrt{A^{\dag}A}$. The trace distance represents a
metric on the space of physical states. It has the important
property that all quantum operations, i.e., all completely
positive and trace preserving (CPT) maps are contractions for this
metric. Given a pair of initial states $\rho_{1,2}(0)$ the rate of
change of the trace distance under the time evolution is defined
by
\begin{equation}\label{sigma}
\sigma(t,\rho_{1,2}(0))=\frac{d}{dt}D(\rho_{1}(t),\rho_{2}(t)).
\end{equation}
A given process is said to be Markovian if for all pairs of
initial states the rate of change of the trace distance is smaller
than zero for all times, i.e., $\sigma(t,\rho_{1,2}(t))\leq0$.
Thus, a process is defined to be non-Markovian if there exists a
pair of initial states $\rho_{1,2}(0)$ and a certain time $t$ at
which the trace distance increases, $\sigma(t,\rho_{1,2}(t))>0$.
As shown in Ref.~\cite{NMmeasure,NMmeasure2} one can construct on the basis
of this definition a measure for non-Markovianity which represents
a functional $\mathcal{N}(\Phi)$ of the corresponding quantum
dynamical map $\Phi$. This measure is defined as the maximum over
all pairs of initial states of the total increase of the
distinguishability during the whole time-evolution:
\begin{equation}\label{N}
\mathcal{N}(\Phi)=\max_{\rho_{1,2}(0)}\int_{\sigma>0}\sigma(t,\rho_{1,2}(0))dt.
\end{equation}

In the following we are going to prove that for the dynamics of a
simplified spin-boson model generated by the generalized memory
kernel master equation~\cite{SabrinaCP} and by the Shabani-Lidar
post-Markovian master equation~\cite{ShabLid} the rate of change
of the distinguishability of any pair of states is always
negative, implying that the measure of non-Markovianity is equal
to zero. Thus, despite the presence of the time-integral over the
past history, the two master equations do not describe any
feedback of information from the environment to the system and are
thus memoryless in this sense. However, it is important to note that
the treated master equations can describe time-dependent uni-directional flow
of information from the system to the environment.

\section{Memory kernel and post-Markovian master equations}

We first present a paradigmatic example of a phenomenological
memory kernel master equation, describing the dynamics of a spin
1/2 interacting with a bosonic reservoir at temperature $T$ under
the rotating-wave approximation,
\begin{equation}\label{MemKerME}
\frac{d\rho(t)}{dt}=\int_{0}^{t}k(t')\mathcal{L}\rho(t-t')dt'.
\end{equation}
Here, $\rho(t)$ represents the reduced density matrix of the spin,
$k(t)$ is a memory kernel function containing information about
the properties of the reservoir, and $\mathcal{L}$ is a Markovian
superoperator. This superoperator is given by
\begin{equation}\begin{split}\label{Liouvillian}
\mathcal{L}\rho=&\frac{\g_{0}(N+1)}{2}(2\si_{-}\rho\si_{+}-\si_{+}\si_{-}\rho-\rho\si_{+}\si_{-})\\
&+\frac{\g_{0}N}{2}(2\si_{+}\rho\si_{-}-\si_{-}\si_{+}\rho-\rho\si_{-}\si_{+}),
\end{split}\end{equation}
with $\g_{0}$ being the phenomenological dissipation constant, $N$
the mean number of excitations of the reservoir, and $\si_{\pm}$
the usual raising and lowering operators of the spin. We consider
a widely used form for the memory kernel function, namely an
exponential function
\begin{equation}\label{memker}
 k(t)=\g e^{-\g t}.
\end{equation}

The memory kernel master equation \eqref{MemKerME} can be solved
using the method of the damping basis \cite{Briegel}. In Ref.
\cite{SabrinaP} the solution for the components of the Bloch
vectors was derived from which one easily obtains the solution
for the spin density matrix $\rho(t)$ corresponding to a generic
initial state. Using the basis $\{\ket{0},\ket{1}\}$ of the
eigenstates of $\sigma_z$ the elements of $\rho(t)$ can be
written as
\begin{equation}\begin{split}\label{densityelements}
\rho_{11}(t)&=u(t)\rho_{11}(0)+v(t)\rho_{00}(0),\\
\rho_{00}(t)&=(1-u(t))\rho_{11}(0)+(1-v(t))\rho_{00}(0),\\
\rho_{10}(t)&=z(t)\rho_{10}(0),
\end{split}\end{equation}
where $u(t),\ v(t)$, and $z(t)$ depend on the damping matrix
$\Lambda={\rm diag}\{\la_{1},\la_{2},\la_{3}\}$ and the
translation vector $\overrightarrow{T}=(T_{1},T_{2},T_{3})$ as
$u(t)=(1+T_{3}+\la_{3})/2$, $v(t)=(1+T_{3}-\la_{3})/2$ and
$z(t)=\la_{1}$. The damping matrix elements and the translation
vector components can in turn be expressed as~\cite{SabrinaCP}
\begin{equation}\begin{split}\label{lambdamemkern}
&\la_{1}=\la_{2}=\xi_{M}(R/2,t),\\
&\la_{3}=\xi_{M}(R,t),\\
&T_{1}=T_{2}=0,\\
&T_{3}=\frac{\xi_{M}(R,t)-1}{2N+1},
\end{split}\end{equation}
where the function $\xi_{M}(R,t)$ is given by
\begin{equation}\begin{split}\label{ximemker}
\xi_{M}(R,t)=&e^{-\g
t/2}\biggl\{\frac{1}{\sqrt{1-4R}}\sinh\biggl[\frac{\g
t}{2}\sqrt{1-4R}\biggl]\\
&+\cosh\biggl[\frac{\g t}{2}\sqrt{1-4R}\biggl]\biggl\},
\end{split}\end{equation}
with $R=\g_{0}(2N+1)/\g$.

An interesting master equation which interpolates between the
generalized measurement interpretation of the Kraus operators and
the continuous measurement interpretation of the Markovian
dynamics is the Shabani-Lidar post-Markovian master equation. The
general form of this master equation is~\cite{ShabLid}
\begin{equation}\label{ShabLidME}
\frac{d\rho}{dt}=\mathcal{L}\int_{0}^{t}k(t')\exp(\mathcal{L}t')\rho(t-t')dt',
\end{equation}
where once more $\rho(t)$ is the density matrix of the reduced
system, $k(t)$ the Shabani-Lidar memory kernel, and $\mathcal{L}$
is the Markovian superoperator. In the following ${\mathcal{L}}$
is taken to be of the form of Eq.~\eqref{Liouvillian}, and the
kernel function is again an exponential function given by
Eq.~\eqref{memker}. The solution for the
density matrix elements of the spin can be written again in the general
form of Eq.~\eqref{densityelements}, where $u(t),\ v(t)$, and  $z(t)$
depend in the same way as before on the damping matrix elements
and the translation vector components~\cite{SabrinaCP}. They in turn have the same
analytic expressions of Eqs.~\eqref{lambdamemkern} with the
exception that $\xi_{M}(R,t)$ has to be replaced by the quantity
$\xi_{P}(R,t)$ which is given by
\begin{equation}\label{xiShabLid}
\begin{split}
\xi_{P}(R,t)=&\exp{\biggl(-\frac{R+1}{2}\g
t}\biggl)\\
&\times\biggl\{\frac{1}{\sqrt{1-r(R)}}\sinh\biggl[\sqrt{1-r(R)}\frac{(R+1)\g
t}{2}\biggl]\\
&+\cosh\biggl[\sqrt{1-r(R)}\frac{(R+1)\g t}{2}\biggl]\biggl\},
\end{split}
\end{equation}
with $r(R)=4R/(R+1)^2$ and $R=\g_{0}(2N+1)/\g$.

In Ref.~\cite{SabrinaCP} the conditions for the positivity and
complete positivity of the dynamical maps associated to the master
equations \eqref{MemKerME} and \eqref{ShabLidME} were studied.
There it was found that $4R\leq1$ is a necessary and sufficient
condition for the positivity of the dynamical map associated to
the memory kernel master equation, while complete positivity is
satisfied only for moderate and high temperatures of the reservoir.
On the other hand, the dynamical map corresponding to the
post-Markovian master equation \eqref{ShabLidME} is always
completely positive. These results are in agreement with
Ref.~\cite{SabrinaP} where it was noticed that the memory kernel
master equation can be derived from the post-Markovian one in the
limit in which the phenomenological dissipation constant $\g_{0}$
is much smaller than the reservoir correlation decay rate $\g$,
suggesting that the post-Markovian master equation is somehow more
fundamental than the former one. Therefore, while for the
post-Markovian Shabani-Lidar master equation we can freely
investigate the non-Markovianity for the whole range of
parameters, in the case of the memory kernel master equation we
are restricted to the conditions $4R\leq1$ and $N\gg1$.

Having constructed the solution of the two master equations we can
now determine the rate of change of the distinguishability given
by Eq.~(\ref{sigma}), which leads to
\begin{equation}\label{sigmaTLS}
 \sigma(t,\rho_{1,2}(0))
 = \frac{a(t)\frac{d}{dt}a(t)+|b(t)|\frac{d}{dt}|b(t)|}{\sqrt{a^2(t)+|b(t)|^2}},
\end{equation}
where $a(t)=\rho_{11}^1(t)-\rho_{11}^2(t)$ and
$b(t)=\rho_{10}^1(t)-\rho_{10}^2(t)$ represent the differences of
the populations and the coherences of the density matrices
$\rho_1(t)$ and $\rho_2(t)$. It is easy to see that these function
are equal to
$a(t)=\la_{3}(\rho_{11}^1(0)-\rho_{11}^2(0))=\la_{3}a_{0}$ and
$b(t)=\la_{1}(\rho_{10}^1(0)-\rho_{10}^2(0))=\la_{1}b_{0}$, with
$\la_{3}=\xi_{M(P)}(R,t)$ and $\la_{1}=\xi_{M(P)}(R/2,t)$ for the
memory kernel and the post-Markovian master equation,
respectively. The derivative of the trace distance can thus be
written as
\begin{equation}\label{sigmafinal}
\sigma(t)=\frac{a_{0}^2\xi(R,t)\frac{d}{dt}\xi(R,t)+|b_{0}|^2\xi(R/2,t)\frac{d}{dt}\xi(R/2,t)}
{\sqrt{a_{0}^2\xi(R,t)^2+|b_{0}|^2\xi(R/2,t)^2}},
\end{equation}
with $\xi(R,t)=\xi_{M(P)}(R,t)$ in the two cases. At this point we
need to study the properties of these two functions.

Let us consider first the memory kernel master equation: Under the
condition $4R\leq1$, $\xi_{M}(R,t)$ is a positive, monotonically
decreasing function which means that $\frac{d}{dt}\xi_{M}(R,t)<0$
and $\frac{d}{dt}\xi_{M}(R/2,t)<0$. Since $a_{0}^2$ and $|b_{0}|^2$
are obviously positive for any pairs of states we thus have
$\sigma(t)\leq0$. We conclude that the rate of change of the
distinguishability of any pair of initial states always decreases
and, hence, the flow of information from the system to the
environment is never inverted during the dynamics and non-Markovian
effects do not appear. Analogously, also in the case of the
post-Markovian master equation $\xi_{P}(R,t)$ is a positive,
monotonically decreasing function such that $\sigma(t)\leq0$. This
implies that also the post-Markovian master equation does not
describe any feedback of information from the environment to the
open system. In contrast to purely Markovian evolution fulfilling
the semigroup property, the dynamics generated by the memory-kernel
and the post-Markovian master equations can be classified as
time-dependent Markovian processes since they do not fulfill the
semigroup property while the information flow is nevertheless
uni-directional from the system to the reservoir. This is
demonstrated by the time-dependence of the decay rates in the
corresponding time-local description given below.

Indeed, both master equations studied here can be written in the time-convolutionless
form
\begin{eqnarray}\label{TCL-1}
\frac{d\rho(t)}{dt}&=&\frac{\gamma_1(t)}{2}(2\si_{-}\rho\si_{+}-\si_{+}\si_{-}\rho-\rho\si_{+}\si_{-})\nonumber \\
&&+\frac{\gamma_2(t)}{2}(2\si_{+}\rho\si_{-}-\si_{-}\si_{+}\rho-\rho\si_{-}\si_{+})\nonumber \\
&&+\frac{\gamma_3(t)}{2}(2\si_{z}\rho\si_{z}-\si_{z}\si_{z}\rho-\rho\si_{z}\si_{z}),
\end{eqnarray}
where
\begin{eqnarray}\label{TCL-2}
\gamma_1(t)&=&-\frac{N+1}{2N+1}\frac{\frac{d}{dt}\xi(R,t)}{\xi(R,t)},\\
\gamma_2(t)&=&-\frac{N}{2N+1}\frac{\frac{d}{dt}\xi(R,t)}{\xi(R,t)},\nonumber\\
\gamma_3(t)&=&\frac{1}{2}\left( \frac{\frac{d}{dt}\xi(R,t)}{2\xi(R,t)}- \frac{\frac{d}{dt}\xi(R/2,t)}{\xi(R/2,t)}\right),\nonumber
\end{eqnarray}
and $\xi(R,t)=\xi_{M(P)}(R,t)$ in the two cases. Thus we see that
the integro-differential master equations given by
Eqs.~(\ref{MemKerME})-(\ref{memker}) and by Eq.~\eqref{ShabLidME}
can be transformed into a form which
is local in time and
does not involve any
time-integration over a memory kernel.

The decay rate $\gamma_3(t)$ in Eq.~(\ref{TCL-2}) is always
negative. It follows that the dynamical map $\Phi$ corresponding to
the master equation is nondivisible~\cite{NMmeasure2}. On the other
hand, we have found above that the process is Markovian. Thus we
have an explicit example of a nondivisible quantum process with zero
measure for non-Markovianity, ${\mathcal{N}}(\Phi)=0$. The existence
of such processes was already conjectured in Ref.~\cite{NMmeasure2}.
Physically this means that the influence of the decay channel with a
negative rate is overcompensated by the effect of the other channels
with positive rates, such that the distinguishability of quantum
states is still monotonically decreasing. We include these
non-divisible processes which have uni-directional information flow
into the class of time-dependent Markovian processes. This class
also includes processes whose decay rates are time-dependent and
positive quantities~\cite{NMmeasure2}. We emphasize that the
time-dependent uni-directional (time-dependent Markovian) processes
and the reversed information flow (non-Markovian) processes have
important fundamental differences, as described recently, e.g., in
Refs.~\cite{NMQJ, NMmeasure,NMmeasure2}.

Going back to the memory kernel master equation (\ref{MemKerME}),
it is also interesting to notice that the non-appearance of memory
effects depends on the restrictions of the range
of parameters imposed by the requirement of positivity. In fact,
when positivity breaks down for $4R>1$, the hyperbolic sine and
cosine of Eq.~\eqref{ximemker} are replaced by trigonometric sine
and cosine. The function $\xi_{M}(R,t)$ then shows damped
oscillations, its derivative has no definite sign, and,
consequently, there can be intervals of time in which the rate of
change of the trace distance $\sigma(t)$ becomes positive implying
that non-Markovian effects appear. We mention that a violation of
the positivity of the dynamical map in phenomenological master
equations was previously studied by Barnett and Stenholm in
Ref.~\cite{Hazard}. There it was shown that the introduction of an
exponential memory kernel function in the dynamics of a damped
harmonic oscillator can lead to blatantly non-physical behavior.

\section{Discussion and Conclusions}
We have applied a recently developed measure for the degree of
non-Markovianity of quantum processes to the dynamical solutions
of a simplified spin-boson model given by two widely used
integro-differential master equations. It has been demonstrated
that, as long as the requirement of the positivity of the
associated dynamical maps is fulfilled, no non-Markovian behavior
occurs, i.e., the measure of non-Markovianity is equal to zero.
This means that the phenomenological memory kernel master
equations considered here are not able to describe a genuine
non-Markovian behavior involving a backflow of information from
the environment to the open system.

Recently, the exact memory kernel master equation for a two-state
system coupled to a zero temperature reservoir has been
constructed~\cite{exact}, showing that in this case the structure
of the master equation given by Eqs.~\eqref{MemKerME} and
\eqref{Liouvillian} is incompatible with a non-perturbative
treatment of the underlying microscopic system-reservoir model.
The perturbation expansion of the exact memory kernel reveals that
in higher orders a new decay channel appears in the superoperator
\eqref{Liouvillian} which is not present in the standard Born
approximation. Thus, while the exact memory kernel master
equation describes correctly all non-Markovian features of the model,
approximation schemes and phenomenological models can lead to strong restrictions in the
treatment of non-Markovianity.

Generally, one might be tempted to think that the introduction of
a memory kernel necessarily leads to a dynamics with
non-Markovianity and memory effects. However, our results
demonstrate that one needs to be cautious when characterizing the physical
properties of open systems only through the mathematical
structure of their equations of motion. The presence of an integral
over the past history in a phenomenological or approximate master
equation does not necessarily guarantee a proper description of
memory effects, namely, the feedback of information from the
environment to the open system. Even though memory kernel master
equations certainly provide a very useful tool for the description
of non-Markovian quantum processes, our results lead to the
following questions: Which of the commonly used phenomenological
or approximate memory kernel master equations are able to
reproduce the key features of non-Markovianity? Can one
formulate general conditions for the structure of the memory
kernels which guarantee the presence of these features in the
dynamics?

\acknowledgments We thank the Magnus Ehrnrooth Foundation, the
Turku University Foundation, the Emil Aaltonen Foundation, the Finnish Cultural Foundation,
the Turku Collegium of Science and Medicine, the National Graduate
School of Modern Optics and Photonics, and the Academy of Finland
(project 133682) for financial support.

\end{document}